\documentclass{vak2024}
\usepackage{journals}
\usepackage[utf8]{inputenc}
\usepackage{url}
\usepackage{hyperref}

\begin{document}
\title{Optical spectroscopy of host-galaxies of intermediate mass black holes: evolution of central black holes}
\titlerunning{Optical spectroscopy of host-galaxies of IMBHs}  
\author{V.~Goradzhanov\inst{1,3}\and I.~Chilingarian\inst{2}\and M.~Demianenko\inst{4}\and I. Katkov\inst{5,6}\and K. Grishin\inst{1, 7}\and V. Toptun\inst{8}\and E. Rubtsov\inst{1}\and D. Gasymov\inst{1}\and I. Kuzmin\inst{1,3}}
\authorrunning{Goradzhanov et al.} 
	%
	%
\institute{Sternberg Astronomical Institute, M.V.Lomonosov Moscow State University, Moscow, Russia,\\
	    \and
	   Center for Astrophysics -- Harvard and Smithsonian (USA),\\
		\and
		Department of Physics, M.V. Lomonosov Moscow State University,\\
		\and
        Max-Planck-Institut für Astronomie, Königstuhl 17, 69117 Heidelberg, Germany\\
        \and
		New York University Abu Dhabi (UAE),\\
		\and
		Center for Astro, Particle, and Planetary Physics, NYU AD (UAE)\\
        \and
        Universit\'e Paris Cit\'e, CNRS(IN2P3), Astroparticule et Cosmologie, F-75013 Paris, France, \\
        \and
        European Southern Observatory, Karl-Schwarzschildstr. 2, 85748, Garching bei M\"unchen, Germany \\}
\abstract{

Intermediate-mass black holes (IMBHs) with masses below ($2 \times 10^5 M_{\odot}$) are pivotal in understanding the origin and growth mechanisms of supermassive black holes (SMBHs) in galactic nuclei. 
This study focuses on the search and detailed analysis of central lightweight black holes in various galaxies. An expanded sample of IMBH candidates was selected from the RCSED optical spectral catalog, followed by refined spectral observations using large telescopes, including the Magellan, SALT, Keck and CMO telescopes. Analyzing over 70 spectra, we obtained accurate virial masses, stellar population parameters, and kinematics. One significant finding includes the detection of a binary black hole system with masses ($1.7 \times 10^5 M_{\odot})$ and $(1.4 \times 10^6 M_{\odot}$). Our results indicate that IMBHs and their low-mass SMBH counterparts do not necessarily co-evolve with their host galaxies, suggesting super-Eddington accretion as a dominant growth mechanism. This research enhances the precision of virial mass estimates and offers new insights into the  $M_{BH} - \sigma_{bulge}$ relation, potentially impacting future high-redshift SMBH observations using next-generation facilities.

\doi{10.26119/VAK2024-ZZZZ}
}

\maketitle

\section{Introduction}

About hundred so-called intermediate-mass black holes (IMBH, $M_{BH} <2*10^{5} M_{\odot}$) have already been discovered. During the next decades, even more IMBHs will be revealed by Laser Interferometer Space Antenna (LISA) \citep{2023LRR....26....2A} and Multi-AO Imaging Camera for Deep Observations (MICADO) \citep{2016SPIE.9908E..1ZD, 2024arXiv240406558D} at Extremely Large Telescope (ELT). At present it is not known exactly the origin and mechanisms of their evolution. Understanding the origin and evolution of IMBHs is extremely important in modern astrophysics, as it will give a key to understanding an even more global problem - the mystery of the origin and growth mechanisms of supermassive black holes (SMBHs) in the nuclei of galaxies. This work is devoted to the search and study of central light-weight black holes in galaxies of different types.

The discovery of quasars in the early Universe (z $> 6.3$, only 750-900 Myr after the beginning of the expansion of the Universe) containing SMBHs with masses on the order of $10^{10}M_{\odot}$ \citep{mortlock11} cannot be explained only by accretion of gas onto stellar-mass black holes. Even if black holes were formed just after the beginning of the expansion of the Universe, by collapse of the nuclei of population III stars, it would take more than 1 billion years for the SMBH to grow, unless the accretion rate significantly exceeds the Eddington limit for a long time.

At the same time, IMBHs are too massive to be formed by the gravitational collapse of a single star, but too light to be considered supermassive. However, there are currently hypotheses about the formation of IMBHs as a result of the collapse of supermassive stars of population III \citep{StarsPopIII}. 

SMBHs are thought to co-evolve with the spheroids of their host galaxies (hereafter host galaxies) \citep{kormendy13}. Thus, there is a so-called “scaling relation” between the black hole mass and the velocity dispersion of the host galaxy.

In recent years, we have expanded the known sample of IMBH candidates and confirmed IMBHs several times over (\cite{IMBH,2023AAS...24230903C, VAK21_VG, VAK21_VT}; by optical variability \cite{VAK21_MD, 2024ASPC..535..283D}). The basis of our galaxy sample is the RCSED optical spectral catalog of galaxies \citep{RCSED}, in which in turn we analyzed the SDSS spectra \citep{SDSS_DR7}. The resolution of SDSS spectra (R~2000) is not suitable for obtaining accurate virial masses, but allows us to make an initial selection of galaxies with low-mass black holes at the center. Having selected IMBH candidates and so-called `light-weight' SMBHs (LWSMBH, $<10^6 M_{\odot}$), we followed them up with spectral observations in the optical range on large telescopes (6.5-m Magellan telescope, 10-m Southern African Large Telescope, 10-m Keck Telescope, 2.5-m telescope of Caucasus Mountain Observatory).

\begin{figure}[h]
\begin{center}
\includegraphics[trim=4cm 1cm 1cm 1cm, clip, height=8.0cm, width=16.0cm]{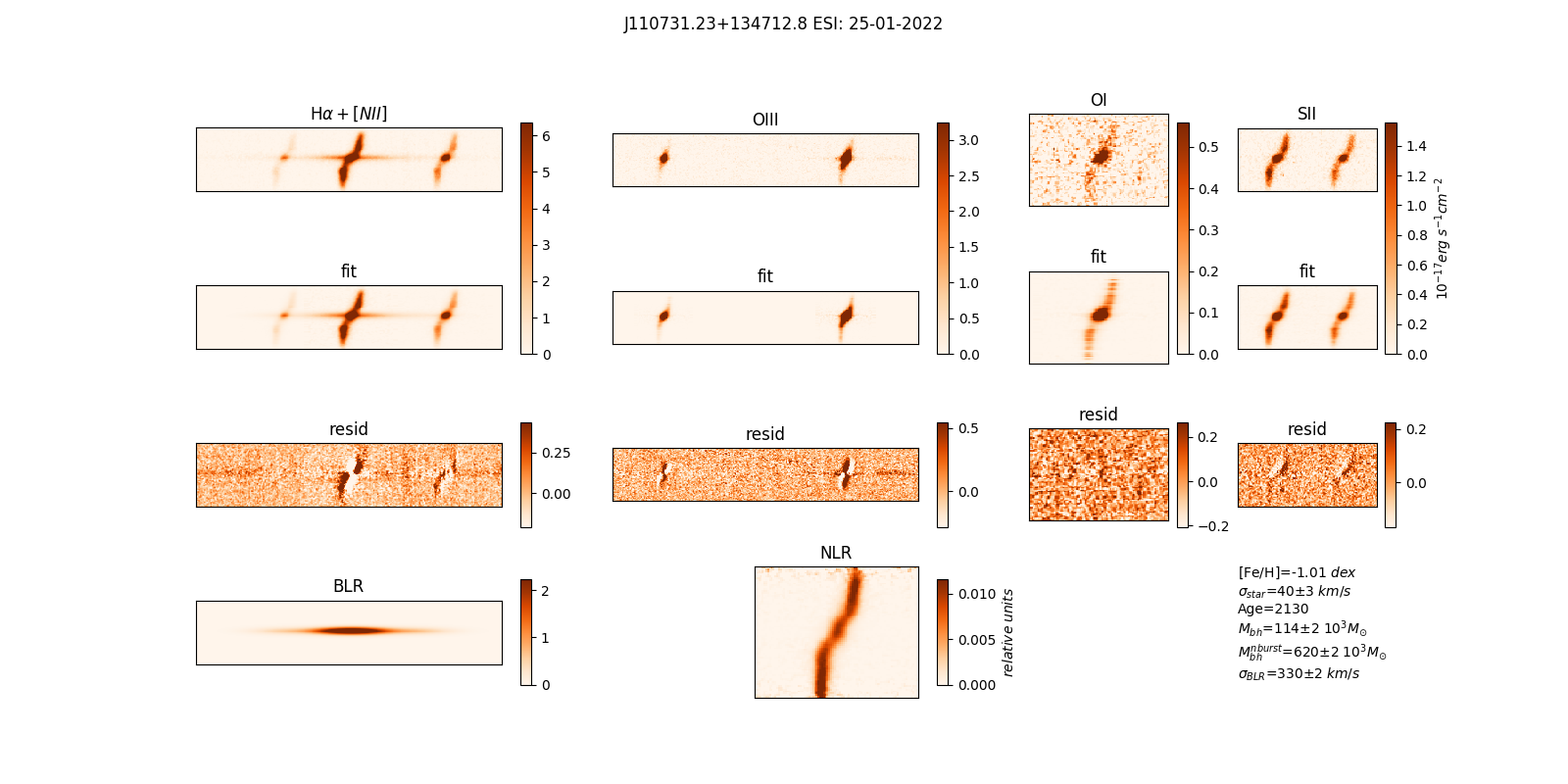}
{\caption{The result of 2d nonparametric reconstruction of the NLR-component profile. The upper line is the emission spectrum of one of the sample galaxies in the selected spectral regions, the second line is the line fit, and the third line is the fit residuals. The fourth line shows the profiles of the parametric BLR component and the non-parametric NLR component. Also on the bottom right the parameters of the BLR component, parameters of the stellar population model - metallicity and age of stars, black hole mass from the NBURSTs decomposition, and from the nonparametric analysis are displayed. \label{fig:2d_fit.png}}}
\end{center}
\end{figure}

\section{Follow-up observations and their analysis}

Spectra of more than 70 galaxies have now been obtained from the MagE (Magellan, echelle), ESI (Keck, echelle), RSS (SALT, long-slit), and TDS (CMO, long-slit) instruments. First, a standardized reduction of our observations was carried out by the pipeline we wrote. The flux calibration of spectra was carried out in order to obtain exact virial masses of black holes. The primary analysis of the spectra consisted in finding the parameters of the stellar population (metallicity, age, kinematics) with our NBURSTs method \citep{NBURSTs}.

Then, the stellar continuum is subtracted from the original spectra and only the emission spectrum is analyzed. A nonparametric decomposition of the $H_{\alpha}$ line profile into BLR and NLR components is performed. The NLR profile reconstruction method simultaneously approximates all strong emission lines: $H_{\beta}, [OIII], [OI], H_{\alpha}, [NII], [SII]$ by a linear combination of a narrow line component having a nonparametric form and a broad Gaussian component in the Balmer lines. The BLR parameters: the velocity dispersion ($\sigma_{BLR}$), the radial velocity of the center of the BLR component, and the parameters h3 and h4 are fit in a nonlinear minimization loop. The parameters psf and the BLR center on the slit are also fit in this loop, since we are dealing with the two-dimensional case. The broad $H\alpha$ flux and width are then used for the virial $M_{BH}$ estimate using the calibration from \citet{reines13}: $M_{BH} = 3.72\cdot10^6 (\mathrm{FWHM}_{H\alpha}/10^3 \mathrm{km/s})^{2.06}\cdot(L_{H\alpha}/10^{42}\mathrm{erg/s})^{0.47}$. The calculated masses were corrected for the slit light losses.

\begin{figure}[htp!]
\begin{center}
\vskip -4mm
\includegraphics[width=0.8\linewidth]{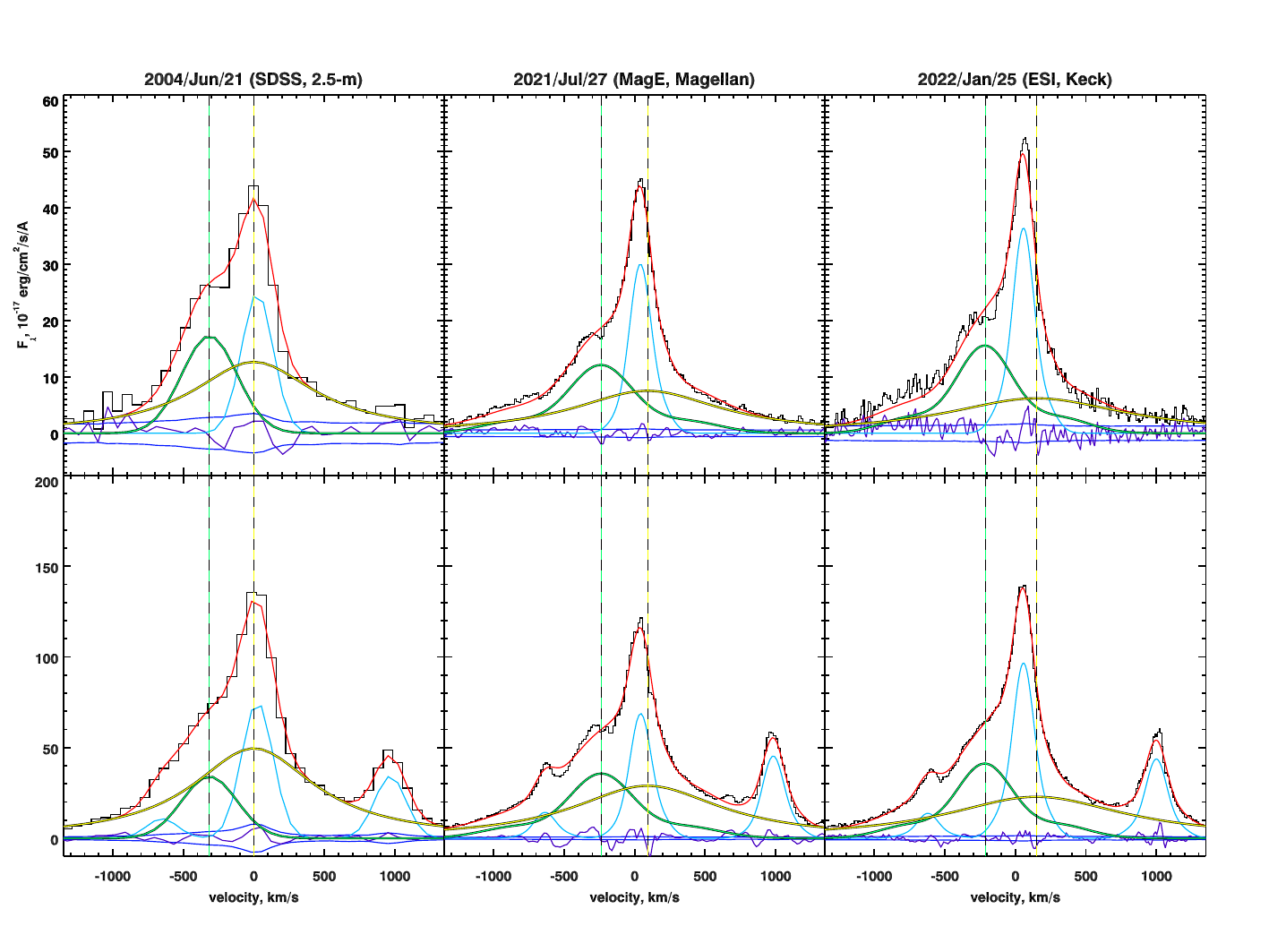}
{\caption{The picture shows spectra of the host galaxy of a binary black hole at different instruments with subtracted stellar continuum. Top panel - spectra (from right to left SDSS, MagE, ESI) of the binary black hole near the $H_\beta$ line, bottom panel - near the $H_\alpha$.}}

\label{fig:1631.pdf}
\end{center}
\end{figure}

\begin{figure}[h]
\begin{center}
\vskip -4mm
\includegraphics[width=0.53\linewidth]{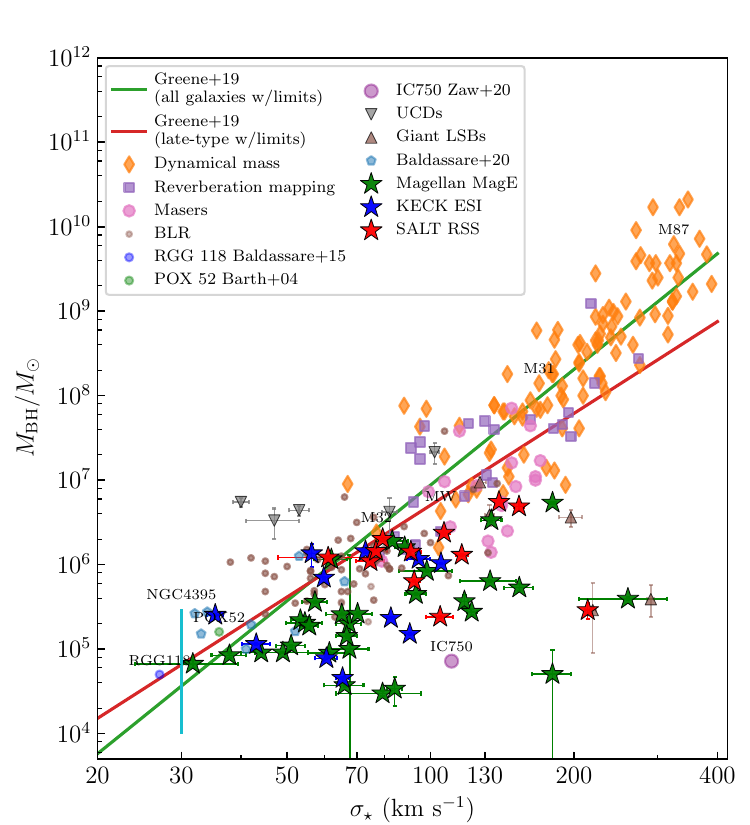}
{\caption{The $M_{BH}-\sigma_*$ relation featuring IMBHs and light-weight SMBHs. Only one IMBH has a mass estimate based on masers \citep{Zaw2020}; the rest of the data points are virial estimates from $H\alpha$. New data from MagE, ESI, and SALT are shown by green, blue, and stars respectively.}

\label{fig:M_sigma.pdf}}
\end{center}
\end{figure}

\section{Evolution of central black holes}

As a result of analyzing one of the spectra obtained at MagE, a binary black hole was detected in one of the galaxies based on the presence of a double broad component in the Balmer line series. The black hole masses are respectively $1.7 \cdot 10^5 M_\odot$ and $1.4 \cdot 10^6 M_\odot$. The two BLR components are 150 km/s apart. After the discovery of this object on MagE, we made observations of this galaxy on other telescopes: Keck and CMO. Currently, this object is under multi-wavelength investigation.

As a result of our analysis the masses of more than 70 IMBH candidates have been refined. The optical spectra (with good spectral resolution) of the currently largest sample of IMBH and LWSMBH have been analyzed: the parameters of stellar populations, the kinematics of stars and gas have been obtained, and the virial mass of central black holes have been measured. The values of virial masses of black holes in comparison with those determined from SDSS spectra have higher accuracy. Due to the higher spectral resolution, it was possible to separate the AGN outflow from the BLR component of the Balmer lines.

Based on the obtained dispersions of velocities in the bulge and virial masses of black holes the dependence $M_{BH} - \sigma_{bulge}$ is constructed. From the results obtained for low-mass black holes, based on the $M_{BH} - \sigma_{bulge}$ relation, we can conclude that IMBHs in the near universe do not appear to evolve with their host galaxies: they grow by accretion, while their hosts grow secularly (even though gas sources may be connected). The scaling relations are preserved, but the dependence of the black hole mass on the velocity dispersion is different. If the same happens at high redshifts, then (super) Eddington accretion is the dominant mechanism for the growth of supermassive black holes at low masses, and we expect to see supermassive black holes with large z in X-rays using the next-generation facilities Athena or Lynx. 
However, there is possibly another explanation for the IMBHs position on the $M_{BH} - \sigma_{bulge}$ relation: scaling relation \citet{reines13} for black hole mass estimation is not valid for small black hole masses. To investigate the reliability of $M_{BH} - FWHM_{H\alpha}$ relations, we started a BLR reverberation campaign using narrow H$\alpha$ filters (Demianenko et al. in prep.) at the CMO of Sternberg Astronomical Institute Moscow State University (SAI MSU).

\textbf{Acknowledgements.} This project is supported by the Basis Foundation Grant 23-2-2-50-1.

\bibliographystyle{aa}
\bibliography{SAO-VAK2024-Article}

\end{document}